\documentclass[twocolumn,superscriptaddress,prb,showpacs]{revtex4}
\usepackage[dvips]{graphicx}
\usepackage[T1]{fontenc}
\usepackage{amsmath}
\usepackage{amssymb}
\usepackage{dcolumn}
\usepackage{bm}

\begin{document}

\preprint{APS/123-QED}

\title{Refinement of the equation of state of tantalum\\}

\author{Agnès Dewaele}
\affiliation{DIF/Département de Physique Théorique et Appliquée, CEA,
BP 12, 91680 Bruy\`eres-le-Ch\^atel, France \\}
\author{Paul Loubeyre}
\affiliation{DIF/Département de Physique Théorique et Appliquée, CEA,
BP 12, 91680 Bruy\`eres-le-Ch\^atel, France \\}
\author{Mohamed Mezouar}
\affiliation{European Synchrotron Radiation Facility, BP 220,
F-38043 Grenoble Cedex, France\\}
\date{\today}

\begin{abstract}
The volume of tantalum versus pressure has been accurately
measured up to 101 GPa by single-crystal x-ray diffraction, with
helium as pressure transmitting medium. Slight deviation from
previous static determinations is observed. Discrepancy with
reduced shock-wave and ultrasonic data supports recent doubts
about the calibration of the ruby pressure scale. Finally, first
principle calculations of the literature show a positive curvature
in $P(V)$ relative to the experimental data, even with a modified
pressure scale.
\end{abstract}

 \maketitle


Ta is a transition metal (Z=73 and the Xe 4f$^{14}$5d$^3$ 6s$^2$
 equilibrium atomic configuration), with a simple bcc
structure stable to at least 174 GPa\cite{Cynn99}. Because of its
fundamental interest and its use as an important technology
material, the equation of state (EoS) of Ta has been studied by
several experimental techniques\cite{Katahara76},\cite{McQueen70},
\cite{Wang02},\cite{Cynn99}, \cite{Hanfland02} and numerous {\it
ab-initio} density functional theory (DFT) electronic structure
calculations\cite{Sorderlin98}, \cite{Wang00}, \cite{Cohen01},
\cite{Boettger01}. However, the EoS of Ta is still not known with
a satisfactory accuracy. On the experimental side, small
differences between static determinations are observed
\cite{Cynn99}, \cite{Hanfland02}, and more important discrepancies
exist among ultrasonic measurements\cite{Katahara76}, shock-wave
results \cite{McQueen70}, \cite{Wang02}, and static determinations
\cite{Cynn99}, \cite{Hanfland02}. On the theoretical side,
systematic deviations are seen between calculations with the two
mostly used approximations of the exchange-correlation energy of
the electrons, and reliable experimental data are needed to
validate one of these approximations\cite{Boettger01}. Here below,
we report accurate EoS data of Ta, based on single-crystal x-ray
diffraction (XRD), up to 101 GPa under quasi-hydrostatic
conditions. We compare these data with previous measurements and
DFT calculations.

Six runs were dedicated to the measurement of accurate volume data
of Ta versus pressure by  synchrotron single crystal x-ray
diffraction, on the ID30 beamline at the ESRF (Grenoble, France).
Small single crystal grains (4 $\mu$m in the maximum dimension),
chosen in a tantalum powder (averaged size 10 $\mu$m, 99.9 \%
purity, Goodfellow product) on the basis of their external shape,
were loaded in membrane diamond anvil cells, with helium as
pressure transmitting medium. We checked by interferometry that
the thickness of  sample chamber was always larger than the
dimension of the crystal. The first two experiments have been
carried out using single crystal energy dispersive XRD and the
other four using angle dispersive monochromatic XRD at 0.3738 \AA.
The same technique had been used to measure with high accuracy the
equation of state of low-Z systems in the 100 GPa
range\cite{Occelli03}. An average of 7 reflections were measured
in each run ; absolute uncertainty in the lattice parameter is at
maximum of 10$^{-3}$. The pressure was estimated from the
luminescence of a small ruby ball (less than 4 $\mu$m in diameter)
and its quasi-hydrostatic calibration\cite{Mao86}. Non-hydrostatic
stresses were observed to be negligible because the shape of the
diffraction spot for a given reflection showed no sign of
deterioration of the crystalline quality of the sample. Except in
one run, when a leak of helium through a diamond fracture caused a
deterioration of the pressurization conditions above 70 GPa, and
the volume of a strained sample, bridged between the diamond tips,
was subsequently measured.


\begin{table}[htb]
\begin{tabular}{cccccccc}
\hline
 $P$ &  V (\AA$^3$) &$P$& V (\AA$^3$) &      $P$       & V (\AA$^3$)      &    $P$     &  V (\AA$^3$)   \\
\hline
10.9  & 17.142&44.2     &15.204     &   9.2   &  17.243 &      52.5   &  14.814         \\
32.0  & 15.717& 54      & 14.761    &   10.6   & 17.140  &     58.7    & 14.554          \\
41.2  & 15.300&    35.3   &15.573      &12.0   & 17.026  &     65.5     &  14.266       \\
11.4  & 17.088 &  52.5    & 14.763     &13.5   & 16.942  &      70      &   14.143     \\
15    &16.815 &   71.9    & 14.047     &15.0   & 16.833  &      90      &   13.449       \\
20.6  & 16.473&  82.5     & 13.697    & 16.1   & 16.751  &       31     &   15.843      \\
25.8  & 16.178&  84.8   & 13.554      & 17.4   & 16.662   &      36.2   &   15.585        \\
31.3  & 15.880&  85.8    & 13.541     & 19.1   & 16.557   &      39.7   &   15.415          \\
36.0  &15.554&     91.3  & 13.379  &    20.8   & 16.449   &       45.6  &    15.135      \\
 41.5 & 15.351&   95.9    &13.225  &     22.4   & 16.353  &       50.2  &    14.919          \\
47    &15.069&  100.9   &13.145  &      24.0   & 16.263   &       55.5  &    14.693      \\
51.4  &14.853&2.3   & 17.838    &       26.2   & 16.129  &       66.5  &    14.257       \\
56.8  &14.566&3.6   &  17.704    &       27.9   & 16.036 &       74.1  &    13.998       \\
60.8  &14.453&5.0   & 17.585    &       30.1   & 15.914  &       79.9  &    13.795       \\
 64.3  &14.42& 6.7 & 17.436 &         1.1  &  17.933   &         85.8&      13.634        \\
 65.7  &14.307 & 7.8 & 17.354 &        0       & 18.033 &             &        \\
 \hline
\end{tabular}
\caption{Atomic volume of Ta measured versus pressure by
single-crystal  XRD in helium pressure transmitting medium.}
 \label{tab:eos}
\end{table}
\begin{figure}[tp]
\includegraphics[height=0.95\linewidth,clip=]{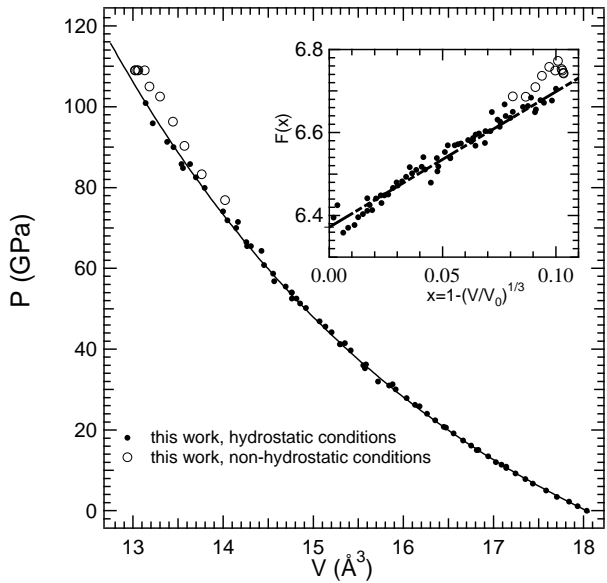}   
\caption{Pressure versus volume data of Ta at 300K. The dots and
circles are respectively the data measured under hydrostatic
conditions and non-hydrostatic conditions. The full line is the
Vinet fit of the data as discussed in the text.
$P=3K_0x^{-2}(1-x)\exp(1.5(K'_0-1)(1-x))$, with
$x=1-(V/V_0)^{1/3}$ and $V_0 =18.033$ \AA$^3$, $K_0=194$ GPa,
$K'_0=3.25$. Inset :
$F(x)=\ln(Px^2/(1-x))=\ln(3K_{0})+\frac{3}{2}(K'_{0}-1)(1-x)$, is
the linearized form of the Vinet function which is used to
determine $K'_0\pm0.1$. \label{fig:eos}}
\end{figure}

 The $P-V$ data points measured
under hydrostatic conditions are presented in Table~\ref{tab:eos}.
They are plotted in Fig. \ref{fig:eos} and compared to the data
obtained with uniaxial stress conditions in the broken diamond
run. It is seen that a non-hydrostatic pressure leads to
overestimate the volume. The ambient pressure atomic volume was
measured equal to 18.033(50) \AA$^3$, which agrees with ambient
$T$ and $P$ literature data (18.065 \AA$^3$, \cite{Alefeld78}).
The uncertainty in the volume determination is smaller than
3.10$^{-3}$, which corresponds to a pressure uncertainty of 1.5
GPa at 100 GPa. The estimated pressure error bars include the
pressure gradient between the ruby ball and the Ta crystal and
range from 0.05 GPa at 1 GPa to 1 GPa at 100 GPa, if the ruby
pressure scale is assumed to be correct. Therefore, the error in
the volume determination, intrinsical to the XRD technique is
mainly responsible of the scattering of the experimental points as
seen in Fig.~\ref{fig:eos}.

To provide the useful physical parameters, namely volume $V_0$,
bulk modulus $K_0$, and its pressure derivative $K'_0$, under
ambient conditions, the $P-V$ data points are generally fitted by
an EoS formulation. The least square fit of the data with an
universal form of the equation of state should ideally give the
same value for $V_0$, $K_0$ and $K'_0$ as the ones measured
independently at zero pressure by XRD for $V_0$ and by ultrasound
propagation for $K_0$ and $K'_0$. We have fitted the data with
three most-used forms of the EoS in the literature: third-order
Birch-Murnaghan EoS \cite{Birch38} (BM), Vinet EoS \cite{Vinet87}
(V) and the EoS proposed by Holzapfel \cite{Holzapfel91} (H). In
this pressure range, very similar values of $K_0$ (between 196 and
199 GPa), and $K'_0$ (between 3.07 and 3.30), are obtained with
these three forms, that give comparable quality of the fits. The
Vinet fit has been chosen to reproduce the present data. In the
case of the Vinet fit, the data can be seen in a linearized form,
as shown in the inset of Fig.~\ref{fig:eos}. It is seen that the
data points above 50 GPa are important to unambiguously constrain
the $K'_0$ value of the slope. The results of the various Vinet
fits of the data (fixing $V_0$, as measured; fixing $V_0$ and
$K_0$ as given by ultrasonic measurements; using the recently
proposed modification of the calibration of the ruby pressure
scale \cite{Holzapfel03}) are compared in table~\ref{tab:EoSparam}
to ultrasonic measurements (the adiabatic to isothermal correction
had been done in ref. \cite{Katahara76}) and to the reduced
Hugoniot data \cite{Wang00}. The $K_0$ value obtained in the
current study, 198 $\pm$ 3.4 GPa, is in correct agreement with the
ultrasonic one, 194 GPa. On the contrary, a smaller $K'_0$ is
obtained, wether $K_0$ is fixed or not to its ultrasonic value.

\begin{table}[htb]
\begin{tabular}{ccccc}
\hline
Study & $V_0$ & $K_0$ & $K'_0$  & Technique, $P$ range, EoS \\
  \hline
current & {\bf 18.033}   &198   &3.07 & XRD, 0-101 GPa, V \\
        &    &(3.4)   &(14) &  \\
current & {\bf 18.033}   &{\bf 194}&3.25 & XRD, 0-101 GPa, V \\
 &    &   &(4) &  \\
Cynn and Yoo\protect\cite{Cynn99} &  {\bf 18.040}  &195&3.4 & XRD, 0-174 GPa, BM \\
Hanfland {\it et al.}\protect\cite{Hanfland02} &  18.191  &207.6&2.85 & XRD, 0-70 GPa, H \\
Katahara {\it et al.} \protect\cite{Katahara76} &    &194&3.83 & US, 0-0.5 GPa \\
Wang {\it et al.} \protect\cite{Wang02} &    &{\bf 194}&3.70 & RSW, 0-150 GPa, V \\
current, & & & & \\
mod $P$ scale\protect\cite{Holzapfel03}  & {\bf 18.033}   &{\bf 194}&3.55 & XRD, 0-105 GPa, V   \\
\hline
\end{tabular}
\caption{Values of $V_0$ (in \AA$^3$), $K_0 $ (in GPa) and $K'_0$
for the EoS of Ta. The current and literature X-Ray diffraction
(XRD) measurements are compared to ultrasonic (US) and reduced
shock-wave (RSW) measurements. For each measurement, the
technique, the pressure range and the form of the EoS (V, BM, H)
are indicated. The parameters that have been fixed in the fitting
procedure are in bold font. Numbers between parenthesis are the
fitting error bars (95 \% confidence interval). The last line
presents the EoS parameters obtained after modification of the
pressure scale\protect\cite{Holzapfel03}. \label{tab:EoSparam} }
\end{table}

\begin{figure}[tp]
\includegraphics[height=.62\linewidth,clip=]{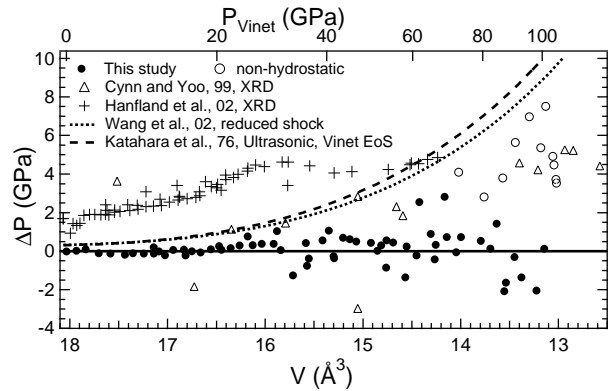}   
\caption{Pressure difference between various experimental
determination of the pressure-volume relationship of Ta. $\Delta
P$ expresses the difference for a given atomic volume, between the
various experimental data and our fitted Vinet EoS, and it is
plotted versus the atomic volume (bottom) and the corresponding
pressure given by the Vinet EoS discussed in the text.
\label{fig:eosExp}}
\end{figure}

In Fig.~\ref{fig:eosExp}, other experimental determinations of the
EoS of Ta are compared through their differences with the Vinet
fit of the present data (with $V_0$=18.033 \AA$^3$, $K_0=194$ GPa,
$K'_0=3.25$). A systematic deviation between the present data and
the ones of two recent XRD measurements is clearly seen. The
numerous low pressure measurements of Hanfland et al. give an
atomic volume at ambient pressure of 18.186 \AA$^3$ (0.8 \% larger
than the present determination). This could be due to a small
contamination by hydrogen (a few percents), that is well-known to
lead to an expansion of the lattice \cite{Alefeld78}. For this
reason, the points plotted in Fig.~\ref{fig:eosExp} appear shifted
from our data. Due to its larger compressibility at low pressure
(Fig.~\ref{fig:eosExp}), the values of $K_0$ fitted on Hanfland
data cannot be reconciled with the ultrasonic value. However, the
compression curve of Hanfland et al. is parallel to the current
one at higher pressure.  Low pressure data points measured by Cynn
and Yoo\cite{Cynn99} exhibit a large scatter. Also, an increasing
deviation between the present EoS data and the data of Cynn is
observed at $P>60$ GPa. This could be ascribed to an uniaxial
stress effect in Cynn experiments\cite{Kenichi01} because these
data are following the trend of our data points in the
non-hydrostatic run (in this study, gold has been used as an x-ray
pressure standard \cite{Heinz84} for the high pressure data
points).

More severe inconsistencies exist between the present XRD and the
ultrasonic measurements or the reduced shock-wave EoS of tantalum.
All XRD studies lead to a smaller $K'_0$ than the ultrasonic one.
Inserting the ultrasonic values \cite{Katahara76} of $K_0$ and
$K'_0$ in the Vinet formulation generates an equation of state
that deviates from our determination at high pressure. As seen in
Fig. 2, the EoS generated from ultrasonic values of $K_0$ and
$K'_0$ is in very good agreement with  the EoS of Ta obtained by
the reduction of shock data to ambient $T$
\cite{McQueen70},\cite{Wang02}. The difference between the XRD EoS
and the reduced shock EoS or the ultrasonic based EoS reaches 12
\% in pressure at 150 GPa. Possible causes of these apparent
inconsistencies need to be examined and error bars estimated. The
error on the reduced Hugoniot data of tantalum \cite{Wang02} is
claimed to be of 5.5 \% at 400 GPa, and less below 200 GPa. This
uncertainty is caused both by uncertainties in particle and shock
velocities measurements, and by the reduction procedure from
Hugoniot temperature to the 300 K. The sum of the error bars in
the volume determination and in the pressure measurements on the
XRD EoS data points amounts to an equivalent 2.5 GPa at maximum at
100 GPa. Consequently, the consideration of cumulative error bars
on the reduced Hugoniot and on the XRD EoS cannot reconcile the
dynamic and static determinations of the EoS of Ta. To reconcile
our XRD data with the reduced Hugoniot data and the ultrasonic
based EoS of tantalum, it must be assumed that the ruby scale
underestimates the pressure by about 10\% at 150 GPa. A similar
correction of the ruby pressure scale has been recently proposed
by Holzapfel \cite{Holzapfel03}. Moreover, a discrepancy between
{\it {ab-initio}} calculations and x-ray measurements of the
pressure-volume relation of diamond has also been ascribed to an
error of the ruby pressure scale of the same magnitude
\cite{Kunc03}. As seen in table~\ref{tab:EoSparam}, by using the
new ruby pressure scale proposed by Holzapfel, consistency is
recovered between the EoS of Ta obtained by static XRD and
shock-wave measurements because similar values of $K_0$, and
$K'_0$ are then obtained. The ultrasonic $K'_0$ value remains
slightly (0.28) larger than the one obtained by fitting of
compression data, which could be due to either experimental
uncertainties or approximations in finite strain EoS formulation.

\begin{figure}[tp]
\includegraphics[height=0.65\linewidth,clip=]{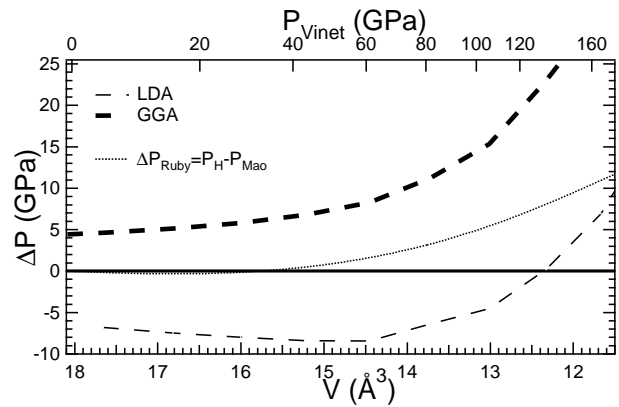}   
\caption{Comparison between DFT calculations and experimental P-V
data points at 300K. The pressure difference, for a given atomic
volume, between the {\it {ab-initio}} pressure and the pressure of
our fitted Vinet EoS is plotted versus volume. GGA and LDA
calculations  \protect\cite{Boettger01} are respectively
represented by bold and thin lines. The pressure change due to
ruby scale scale of Holzapfel \protect\cite{Holzapfel03} is
indicated as dotted line \ \label{fig:abinit}}
\end{figure}

The {\it {ab-initio}} calculations of the EoS of Ta has been
produced in four recent studies \cite{Sorderlin98},
\cite{Wang00},\cite{Boettger01}, \cite{Cohen01} using
density-functional theory. In principle, this method involves
approximations only in the exchange and correlation energy
calculation. Two forms are currently used, namely the Local
Density Approximation (LDA) and the Generalized Gradient
Approximation (GGA). The validity of these approximations is in
practice established by their ability to reproduce experimental
results. Uncertainties might also arise consequently to the choice
of the computational method and of its inherent assumptions, but
they should be small, as it is seen by comparing three recent GGA
EoS \cite{Wang00},\cite{Boettger01}, \cite{Cohen01}. These EoS
agree within 3 GPa at 150 GPa, although they are based on
different computational methods. Consequently, for legibility in
Fig. 3, the LDA and GGA EoS are represented by the calculation of
Boettger \cite{Boettger01} alone, and they are compared to the
experimental data through the pressure difference at a given
volume. It is observed here, as often, that the ambient pressure
volume is underestimated by LDA calculations and overestimated by
GGA calculations. What seems to be more interesting here is that
all DFT calculations show a similar positive curvature relative to
the experimental data. The higher compressibility at high pressure
of the experimental EoS compared to the DFT calculations cannot be
entirely due to the possible error of the ruby pressure scale.  In
fact, if the new calibration proposed by
Holzapfel\cite{Holzapfel03}, that reconciles all the experimental
determinations as discussed above, is used, the larger
experimental compressibility still exists, as shown in
Fig.~\ref{fig:abinit}, for pressures higher than 80 GPa.

In summary, the present work demonstrates that an accurate
determination of the volume of metals under hydrostatic conditions
in the 100 GPa range can now be achieved by synchrotron
single-crystal XRD with helium as pressure transmitting medium.
Consistency with the determination of the EoS of Ta by ultrasonic
or shock-wave methods can be recovered by using a  calibration of
the ruby scale recently proposed. This highlights and estimates
the present uncertainty of the ruby pressure calibration. Similar
EoS measurements under good hydrostatic conditions up to the Mbar
range should now be performed on reference metals to address two
important issues pointed out in the present work: first, to reduce
the uncertainty of the ruby pressure scale, by a comparison of
these metal EoS data with the EoS deduced from high-temperature
shock-wave, considered as primary standards\cite{Mao86}; second,
to investigate if a higher experimental compressibility than the
calculated DFT one is a systematic trend in metals and if it has a
Z dependence.

\begin{acknowledgments}
We acknowledge the European Synchrotron Radiation Facility for
provision of synchrotron radiation facilities and we would like to
thank Anne-Claire Dhaussy and Michaël Berhanu for assistance in
using beamline ID30. We thank Michael Hanfland for providing the
detail of its $P-V$ data points. We are grateful to the two
referees for their helpful comments on the manuscript.
\end{acknowledgments}

\bibliographystyle{apsrev}

\end{document}